\documentclass[12pt]{article}
\usepackage[dvipdfmx]{graphicx}

\usepackage{color}
\usepackage{url}

\newcommand\pubnumber{CIPANP2018-Nagai}
\newcommand\pubdate{September 30, 2018}

\def\CUBoulder{Department of Physics\\
University of Colorado Boulder, CO 80302, USA}
\def\collaboration{\footnote{For the NA61/SHINE collaboration}}

\def\Title#1{\begin{center} {\Large #1 } \end{center}}
\def\Author#1{\begin{center}{ \sc #1} \end{center}}
\def\Address#1{\begin{center}{ \it #1} \end{center}}

\newcommand\pubblock{\rightline{\begin{tabular}{l} \pubnumber\\
         \pubdate  \end{tabular}}}
\newenvironment{Abstract}{\begin{quotation}  }{\end{quotation}}
\newenvironment{Presented}{\begin{quotation} \begin{center} 
             PRESENTED AT\end{center}\bigskip 
      \begin{center}\begin{large}}{\end{large}\end{center} \end{quotation}}
\def\Acknowledgements{\bigskip  \bigskip \begin{center} \begin{large}
             \bf ACKNOWLEDGEMENTS \end{large}\end{center}}

\def\beq{\begin{equation}}
\def\eeq#1{\label{#1}\end{equation}}
\def\eeqn{\end{equation}}

\def\beqa{\begin{eqnarray}}
\def\eeqa#1{\label{#1}\end{eqnarray}}
\def\eeqan{\end{eqnarray}}

\let\bar=\overbar

\def\Dslash{\not{\hbox{\kern-4pt $D$}}}
\def\dslash{\not{\hbox{\kern-2pt $\del$}}}

\def\msb{{\bar{\ssstyle M \kern -1pt S}}}

\begin{document}
\begin{titlepage}
\pubblock

\vfill
\Title{Hadron Production Measurements for Long-Baseline Neutrino}
\Title{Experiments with NA61/SHINE}
\vfill
\Author{ Yoshikazu Nagai \collaboration}
\Address{\CUBoulder}
\vfill
\begin{Abstract}
A precise prediction of the neutrino flux is a key ingredient for achieving the physics goals of long-baseline neutrino experiments.
In modern accelerator-based neutrino experiments, neutrino beams are created from the decays of secondary hadrons produced in hadron-nucleus interactions.
Hadron production is the leading systematic uncertainty source on the neutrino flux prediction; therefore, its precise measurement is essential.

The NA61/SHINE is a fixed-target experiment at the CERN Super Proton Synchrotron, which studies hadron production in hadron-nucleus and nucleus-nucleus collisions for various physics goals.
For neutrino physics, light hadron beams (protons, pions, and kaons) are collided with a light nuclear target (carbon, aluminum, and beryllium) and spectra of outgoing hadrons are measured.

These proceedings will review the recent results and ongoing hadron production measurements with NA61/SHINE for the precise neutrino flux predictions in the T2K and
Fermilab long-baseline neutrino experiments.
It will also discuss the prospect for future hadron production measurements with NA61/SHINE beyond 2020,
after the Long Shutdown 2 of the accelerator complex at CERN.
\end{Abstract}

\begin{Presented}
Thirteenth Conference on the Intersections of Particle and Nuclear Physics (CIPANP2018)\\
Palm Springs, CA, USA,  May 29 - June 3, 2018
\end{Presented}
\end{titlepage}
\def\thefootnote{\fnsymbol{footnote}}
\setcounter{footnote}{0}
%


\section{Introduction}

In this section, 
motivation of hadron production measurements,
types of targets for measurements,
and the NA61/SHINE experimental apparatus will be discussed.

\subsection{Hadron Production}

In modern long-baseline neutrino oscillation experiments, neutrino beams are created by colliding protons with a light nuclear target, such as carbon or beryllium.
Secondary hadrons are produced via primary interactions of beam protons and their decays contribute to the neutrino flux.
Neutrinos mostly come from decays of charged pions, while kaon contributions to the flux are getting larger at higher energy region.
In addition, a significant fraction of neutrinos come from re-interactions of secondary hadrons (pions, kaons, protons, and so on)
in target, aluminum magnetic horn, or with other beamline materials.
Therefore, precise knowledge on hadron production for both primary and secondary interactions is necessary.

\begin{figure}[htb]
\centering
\includegraphics[height=53mm]{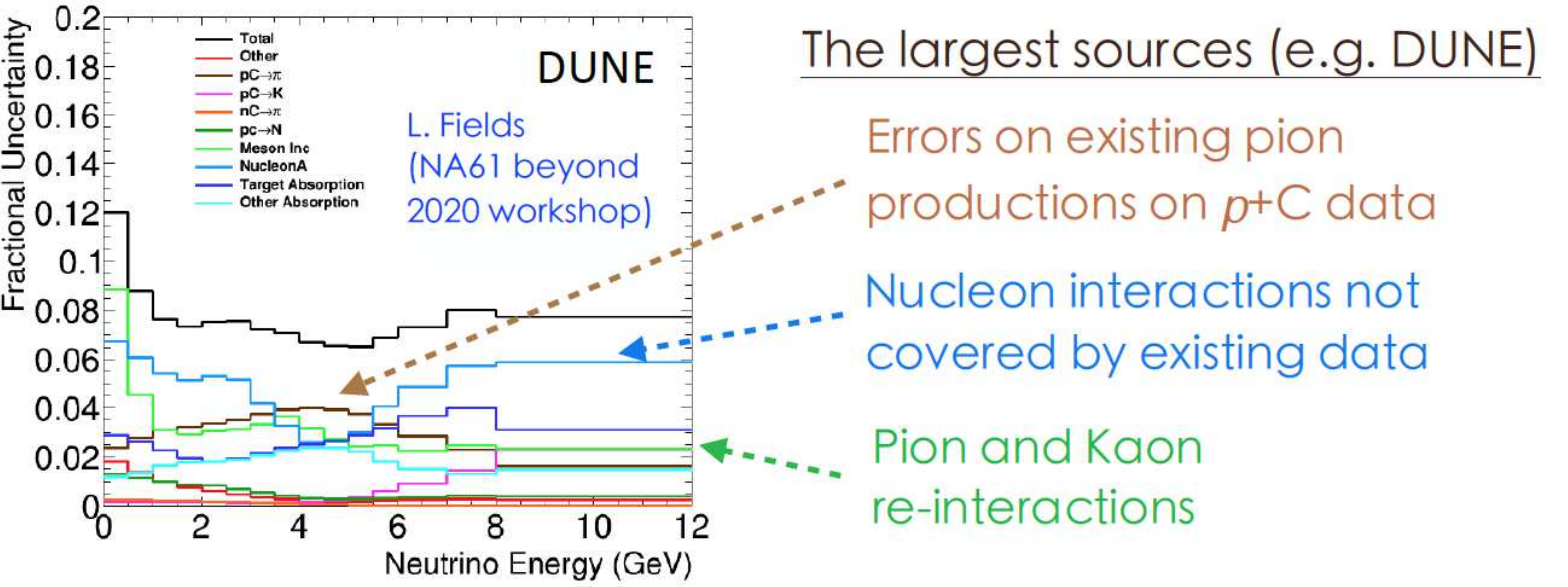}
\caption{Flux uncertainty prediction of the DUNE experiment~\cite{Laura-NA61Beyond2020}.}
\label{Fig:DuneFlux}
\end{figure}

In long-baseline neutrino oscillation analyses, number of observed neutrino events at the near and far detectors can be written as proportional to the neutrino flux and cross-section:
\begin{eqnarray}
N_{ND}&\propto&\int \Phi_{ND} \cdot \sigma \,\, dE_\nu  \nonumber \\ 
N_{FD}&\propto&\int \Phi_{FD} \cdot \sigma \cdot P_{osc} \,\, dE_\nu  \nonumber  \\
            &\propto&\int R_{\frac{FD}{ND}} \cdot \Phi_{ND} \cdot \sigma \cdot P_{osc} \,\, dE_\nu  \nonumber
\end{eqnarray}
where $\Phi_{ND(FD)}$, $\sigma$, $P_{osc}$, and $R_{\frac{FD}{ND}}$ denote neutrino flux at the near (far) detector,   neutrino cross-section on target materials, neutrino oscillation probability, and far to near neutrino flux ratio, respectively. 
Here, near detectors can not fully constrain flux at the far site and large uncertainty remains.
For instance at the T2K experiment, to achieve desired sensitivity for the observation of CP violation,
total systematic uncertainty needs to be below 4\%.
On the contrary, current T2K has about 6\% total uncertainty on the $\nu_e$ appearance channel including 3\% uncertainty on the $\Phi \cdot \sigma$ term.
Similar situation will be expected for the future Hyper-K and future DUNE experiments.
An example of flux uncertainty for the DUNE experiment is shown in Figure~\ref{Fig:DuneFlux}. 

In addition, neutrino cross-section and other near detector measurements rely heavily on a precise flux prediction.
For instance, the flux prediction of the MINER$\nu$A experiment in the NuMI beamline at Fermilab relies on former hadron production measurements~\cite{Aliaga:2016oaz}, however, its coverage is not enough and they assign large uncertainties.
Therefore, hadron production is the leading source of uncertainty and its reduction is desired.

\subsection{Thin and Replica Target Measurement}

There are two types of measurements to constrain hadron production uncertainty: thin and replica targets.

\begin{figure}[htb]
\centering
\includegraphics[height=50mm]{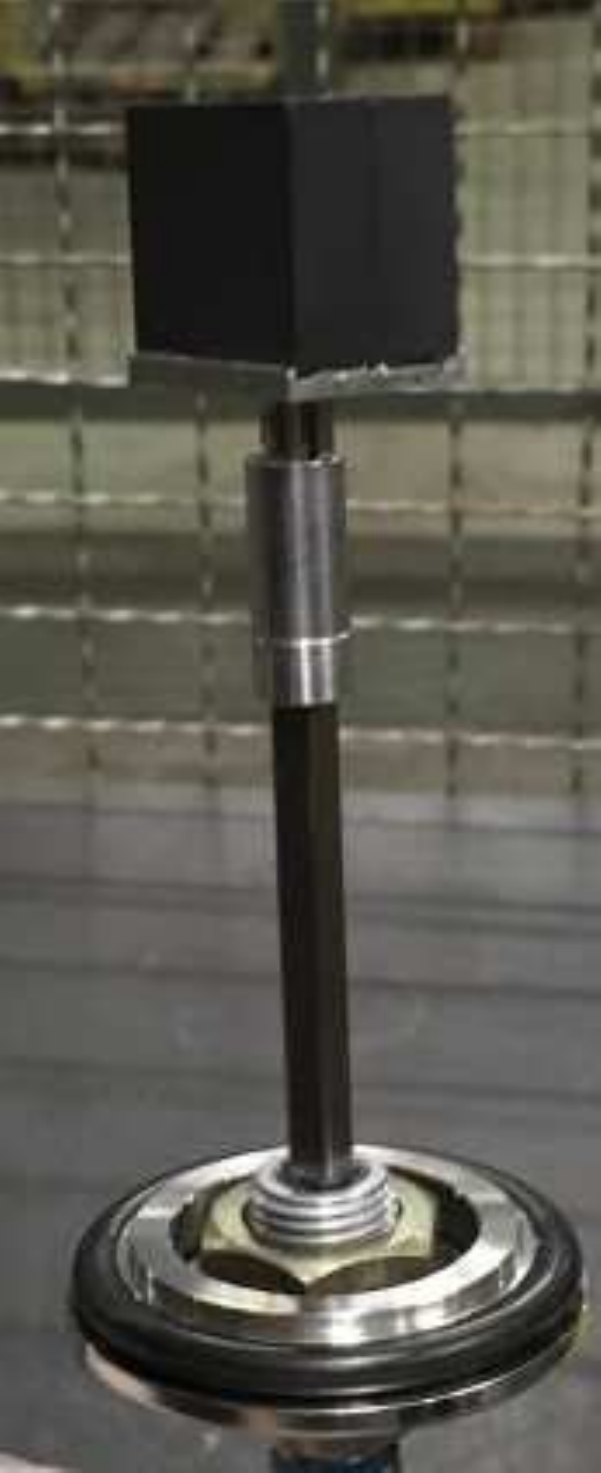} \hspace{5mm}
\includegraphics[height=50mm]{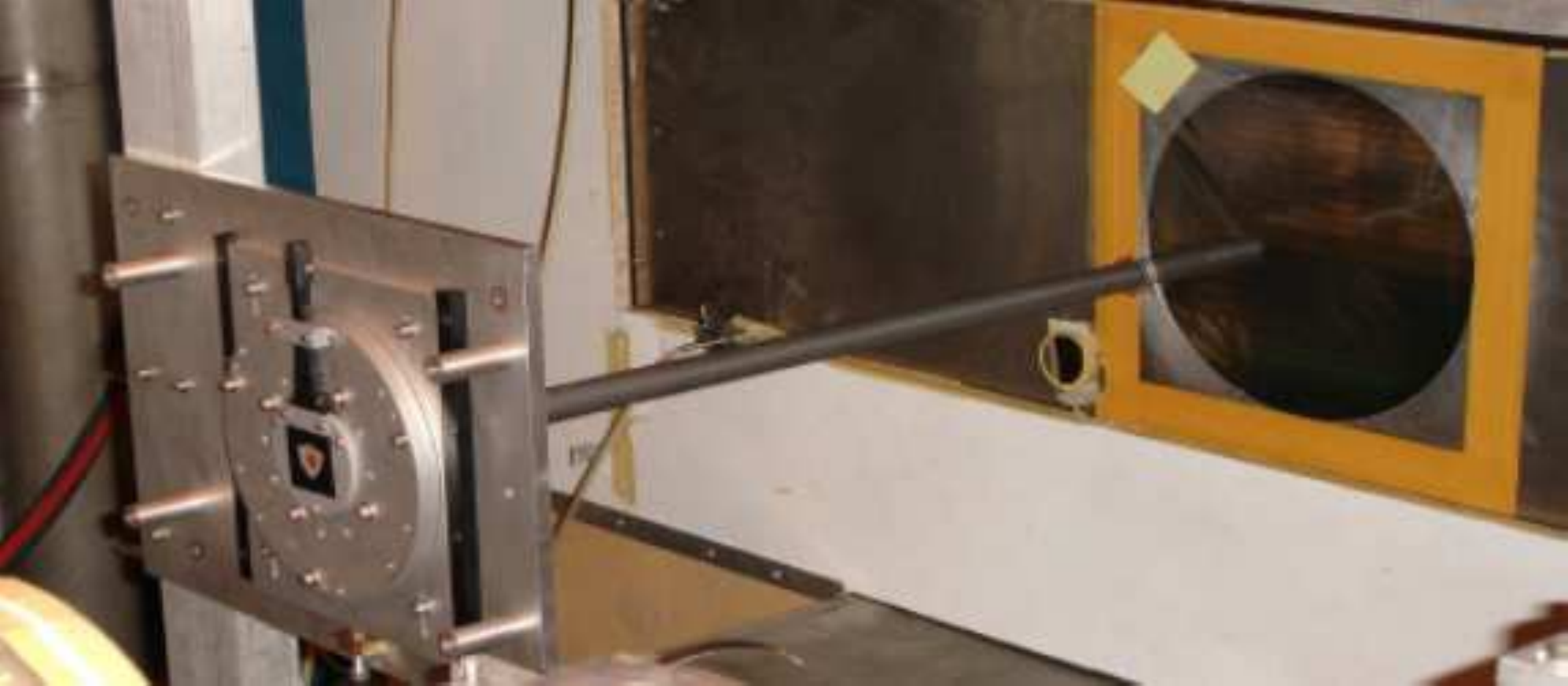}
\caption{(Left) A 1.5cm long thin graphite target ($\sim$3\% of $\lambda$). (Right) A 90cm long T2K replica graphite target ($\sim$1.9 $\lambda$).
}
\label{Fig:target}
\end{figure}

Thin targets (a few \% of proton-nuclear interaction length, $\lambda$) are used to study primary and secondary interactions.
Proton ($p$), pion ($\pi^\pm$), and kaon ($K^\pm$) beams strike nuclear targets (such as carbon, beryllium, aluminum)
and secondary hadrons are produced.
To constrain the uncertainty on production, measurements of total inelastic and production cross-sections as well as differential cross-sections ($\frac{d^2\sigma}{dpd\theta}$) are made.
A typical thin graphite target is shown in Figure~\ref{Fig:target} (Left).


Replica targets (often also called thick targets) are used to study hadrons exiting target and to measure
production cross-section via beam survival probability:
$P_{\textrm{\footnotesize{survival}}} = e^{-Ln\sigma_{\textrm{\scriptsize{prod}}}}$.
An example of a T2K replica graphite target is shown in Fig~\ref{Fig:target} (Right).

\subsection{The NA61/SHINE experiment}

\begin{figure}[htb]
\centering
\includegraphics[height=49mm]{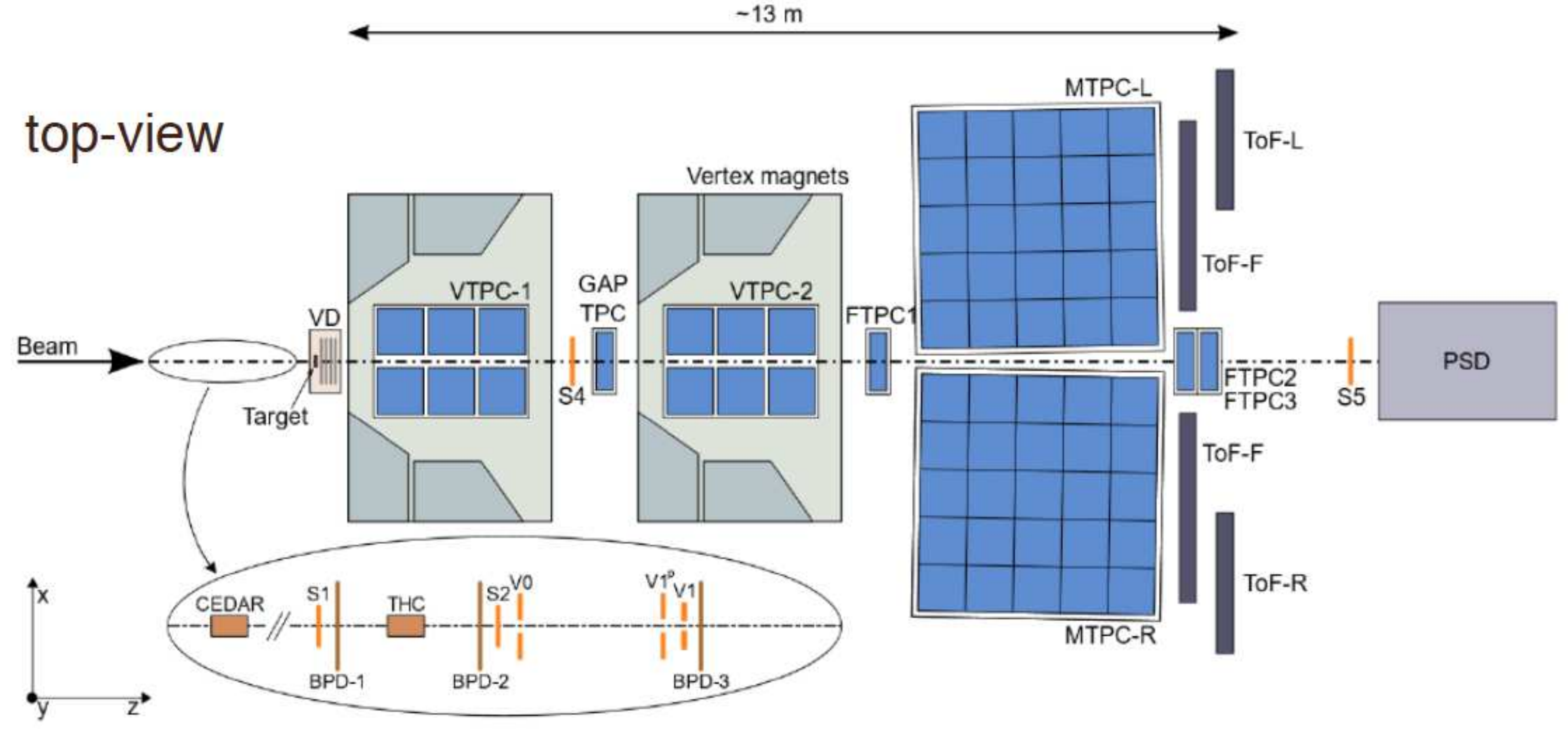} \hspace{2mm}
\includegraphics[height=49mm]{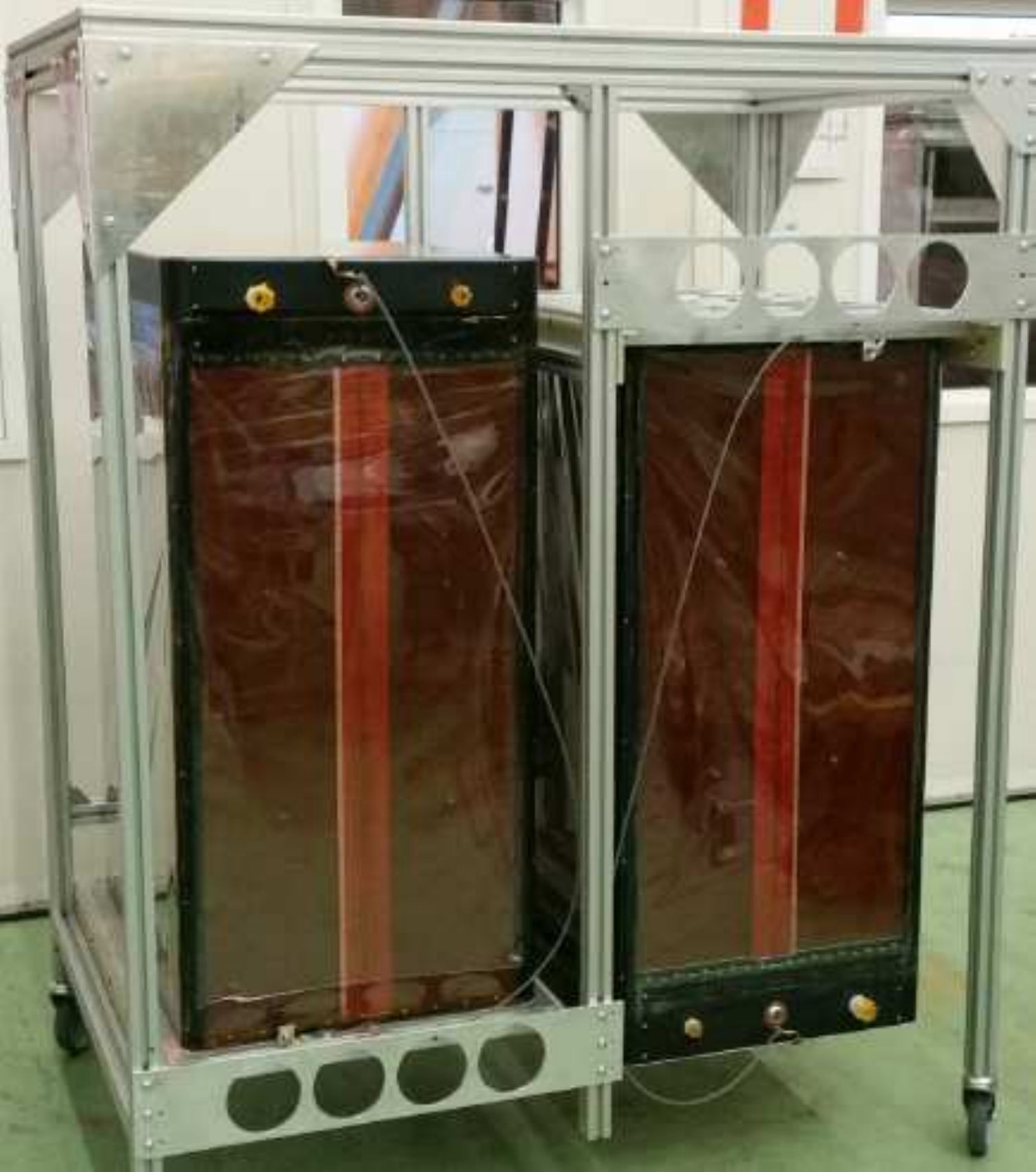}
\caption{(Left) The schematic top-view layout of the NA61/SHINE experiment. (Right) Constructed FTPC 2 and FTPC 3 chambers.
}
\label{Fig:NA61}
\end{figure}

\begin{figure}[htb]
\centering
\includegraphics[height=55mm]{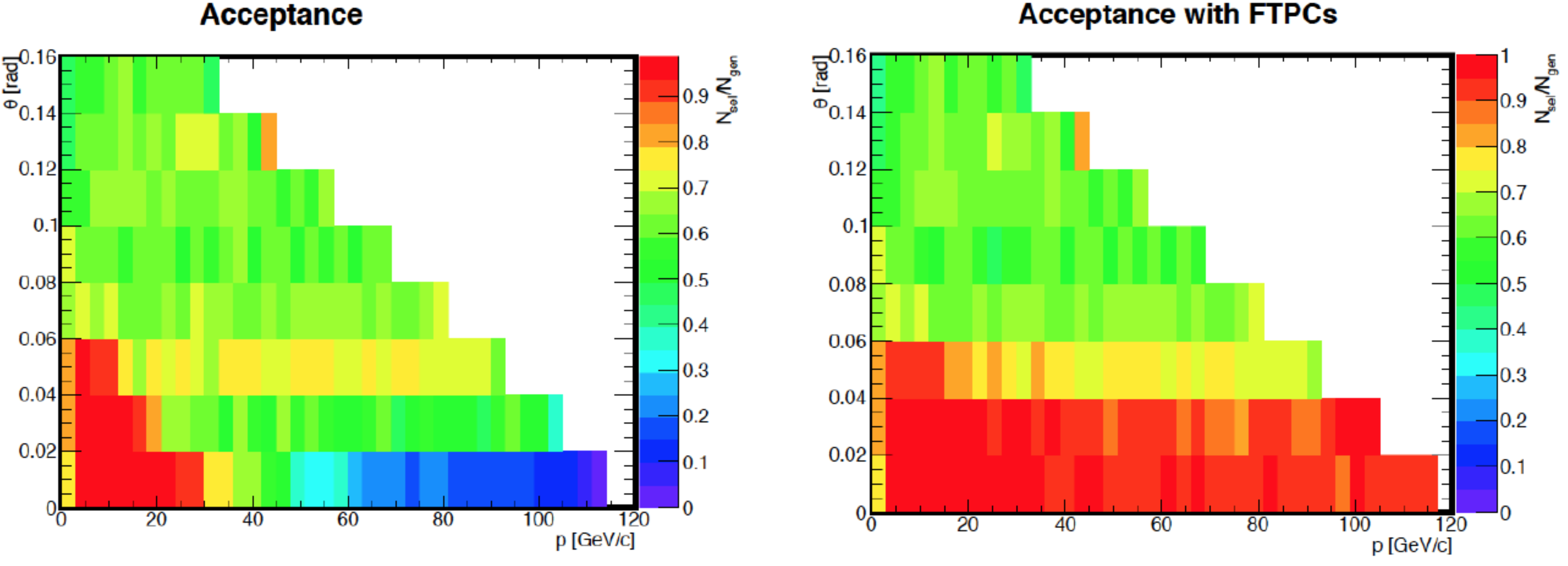}
\caption{ NA61/SHINE acceptance as a function of production angle and momentum.  (Left) without FTPCs. (Right) with FTPCs.
}
\label{Fig:FTPCacceptance}
\end{figure}

The NA61/SPS Heavy Ion and Neutrino Experiment (NA61/SHINE)
is a fixed-target experiment at the CERN SPS, which studies hadron production in hadron-nucleus and nucleus-nucleus collisions for various physics goals.
The NA61/SHINE apparatus is a large acceptance spectrometer for charged particles.
Figure~\ref{Fig:NA61} (Left) shows the NA61/SHINE experimental setup. 
Main tracking detectors are four large TPCs (VTPC-1, VTPC-2, MTPC-L, MTPC-R), where
two of them sit inside the super-conducting dipole magnets with a combined maximum bending power of 9\,Tm,
and other two are located downstream of magnets symmetrically with respect to the beamline.
To fill the gap on the beam direction, GapTPC is positioned along the beam axis between two VTPCs.
Moreover, forward TPCs (FTPC 1, 2, 3) have been installed in 2017 (Figure~\ref{Fig:NA61} (Right)) and huge improvement on the forward acceptance has been achieved (Figure~\ref{Fig:FTPCacceptance}).
This facility update is particularly important to understand forward proton and pion productions for NuMI and LBNF beamlines.
These TPCs provides good momentum reconstruction and particle identification capabilities. 
Scintillator-based time of flight detectors (ToF) are located downstream of TPCs, which give complementary particle identification capability especially for the region where the Bethe-Bloch dE/dx curves overlap.
The Projectile Spectator Detector (PSD), a forward hadron calorimeter, sits downstream of the ToF system.

\section{Results and Plans}

In this section, NA61/SHINE latest results and ongoing measurements for the T2K and Fermilab neutrino program will be presented.

\subsection{Measurements for the T2K flux prediction} 

NA61/SHINE performed thin and replica target measurements to reduce the uncertainty on the T2K flux prediction.

\begin{figure}[htb]
\centering
\includegraphics[height=52mm]{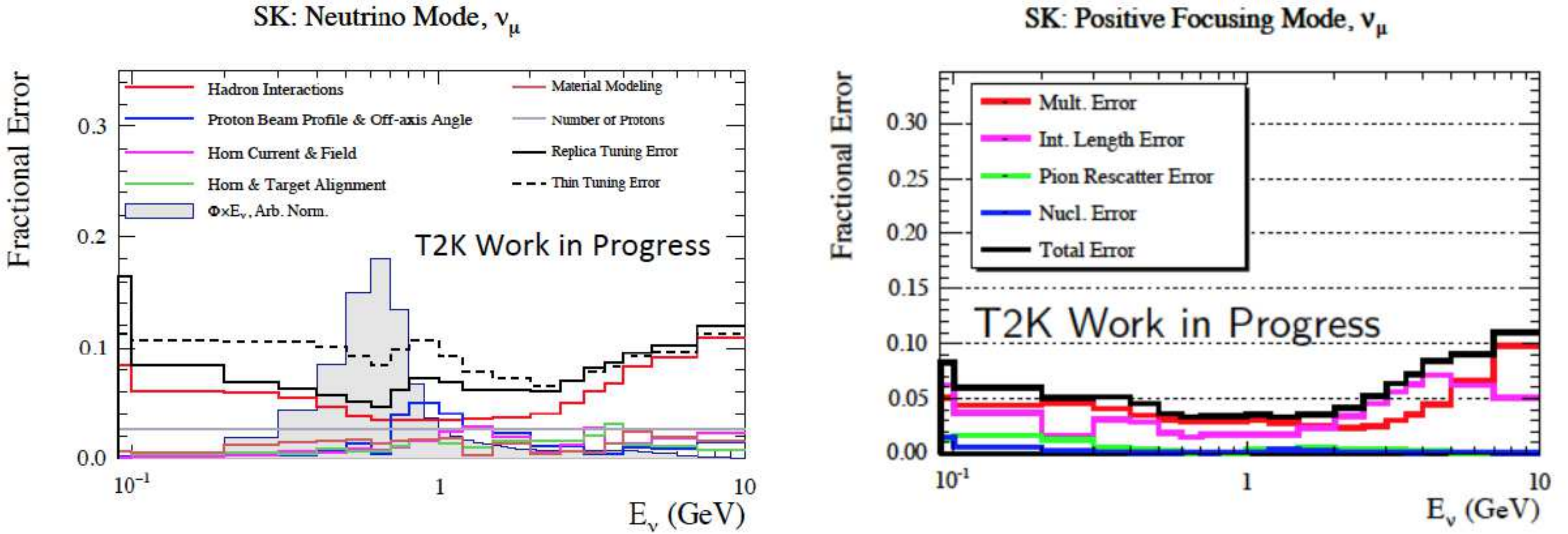}
\caption{ T2K flux uncertainty at the far detector. (Left) Black dotted and solid lines are flux uncertainty tuned with thin target measurements and replica target measurement ($\pi^\pm$ yields only), respectively.
(Right) Expected flux uncertainty improvement with the latest replica target measurements.
}
\label{Fig:T2Ktuning}
\end{figure}

To constrain primary proton beam interactions on carbon target,
proton beam at 31 GeV/$c$ was shot on a thin carbon target (2cm long, $\sim$4\%\,$\lambda$, denote as $p$+C).
These interactions were used for measurements of total cross-sections and spectra measurements of $\pi^\pm$, $p$, $K^+$,$K^0_S$,  and $\Lambda^0$.
NA61/SHINE has published these results in \cite{Abgrall:2011ae, Abgrall:2011ts, Abgrall:2013wda, Abgrall:2015hmv}.
These thin target measurements successfully improved flux uncertainty down to 10\% (Figure~\ref{Fig:T2Ktuning} (Left), black dotted line).

\begin{figure}[!htb]
\centering
\includegraphics[height=64mm]{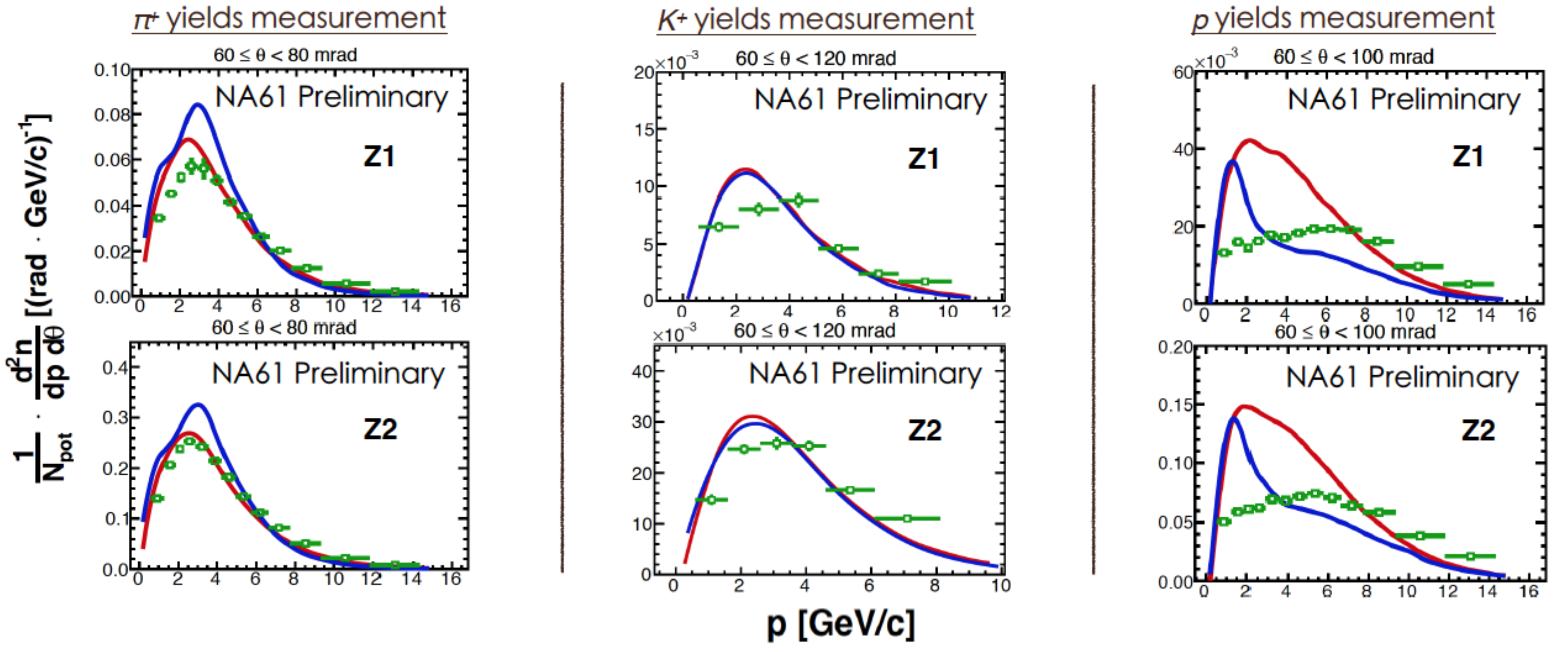}
\caption{ Yield of hadrons T2K replica target. From left to right, $\pi^+$, $K^+$, and proton yields are shown, respectively. NA61/SHINE data points (green points) are compared with predictions of NuBeam (red line) and QGSP\_BERT (blue line) predictions with Geant4 version 10.03. 
}
\label{Fig:T2KReplicaYields}
\end{figure}

To further constrain flux uncertainty,
NA61/SHINE measured $\pi^\pm$ spectra using proton beam at 31 GeV/$c$ on the T2K replica target~\cite{Abgrall:2012pp, Abgrall:2016jif} (Figure~\ref{Fig:T2Ktuning} (Left), black solid line).
Recently, NA61/SHINE has performed improved measurements with five times larger statistics~\cite{Berns:2018tap}.
Yields of hadrons ($\pi^\pm$, $p$, $K^\pm$) exiting the surface of the T2K replica target has been measured and compared with several model predictions. 
Figure~\ref{Fig:T2Ktuning} (Right) shows expected improvement on the flux uncertainty and
Figure~\ref{Fig:T2KReplicaYields} shows an example of observed differential yields of hadrons. 


\subsection{Measurements for the Fermilab neutrino program}

Measurements for the Fermilab neutrino beamlines, NuMI and LBNF, are ongoing with NA61/SHINE.

\begin{figure}[htb]
\centering
\includegraphics[height=54mm]{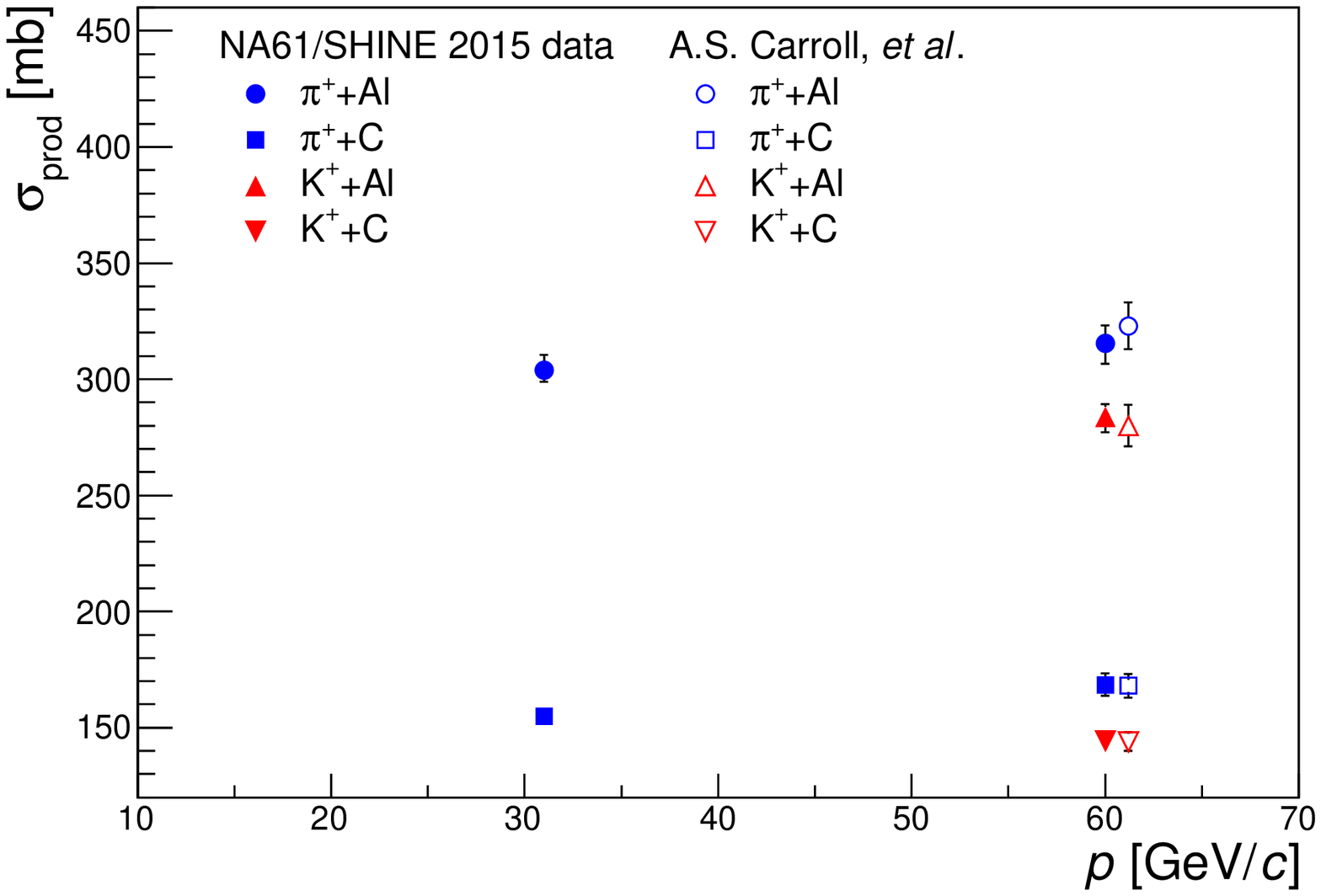} \hspace{-9mm}
\includegraphics[height=54mm]{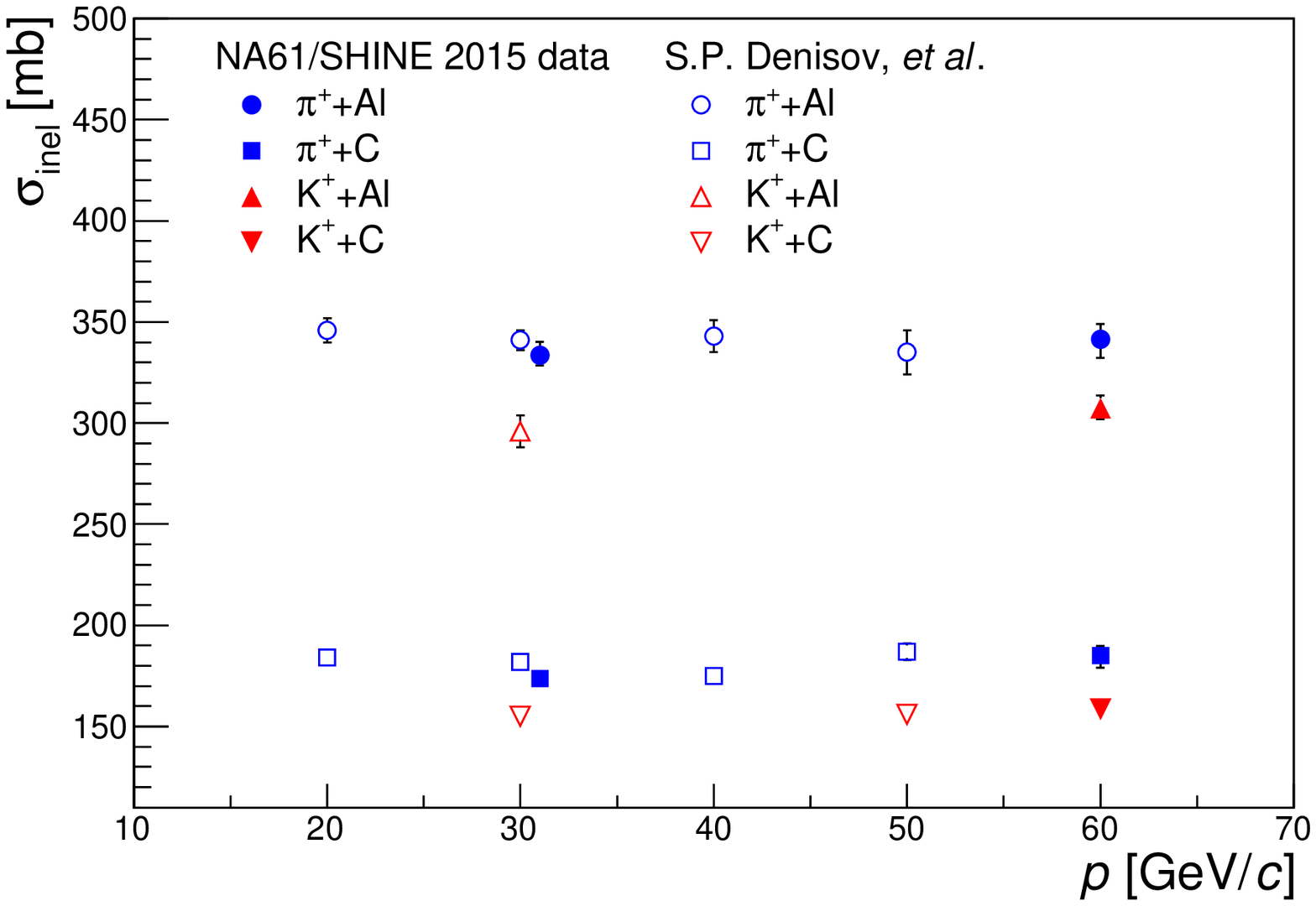}
\caption{ Results of cross-section measurements with the 2015 NA61/SHINE data set. (Left) Production cross-section. (Right) Inelastic cross-section.
}
\label{Fig:NA61totalSigma}
\end{figure}

In 2015, superconducting magnets were not operational and spectral measurements were not possible.
Therefore, experimental setup was optimized to measure inelastic and production cross-sections,
which they are defined in terms of the inelastic $\sigma_\mathrm{inel}$ and coherent elastic $\sigma_\mathrm{el}$ cross-sections:
\begin{eqnarray}
\sigma_\mathrm{total} = \sigma_\mathrm{inel} + \sigma_\mathrm{el}. \nonumber 
\end{eqnarray}
The inelastic cross-section $\sigma_\mathrm{inel}$ is defined as the sum of all processes due to strong interactions
other than coherent elastic scattering.
In case  new hadrons are produced, process is taken into account the production $\sigma_\mathrm{prod}$ cross-section.
Considering that the quasi-elastic scattering is a part of the inelastic scattering, the production cross-section $\sigma_\mathrm{prod}$ can be difined in terms of the quasi-elastic scattering $\sigma_\mathrm{qe}$:
\begin{eqnarray}
\sigma_\mathrm{prod} = \sigma_\mathrm{inel} - \sigma_\mathrm{qe}. \nonumber
\end{eqnarray}
Measurements of six different interactions on thin carbon and aluminum targets using $\pi^+$ and $K^+$ beams have been recently published~\cite{Aduszkiewicz:2018uts}.
Results are summarized in Figure~\ref{Fig:NA61totalSigma}.
Precision of these measurements are typically 2\,--\,3\%, while current NuMI beam for the MINER$\nu$A experiment assumes 5\% (10\,--\,30\%) for pions (kaons)~\cite{Aliaga:2016oaz}.
Thus, these measurements will significantly reduce the uncertainty on the neutrino flux prediction in NuMI. 

\begin{figure}[htb]
\centering
\includegraphics[height=70mm]{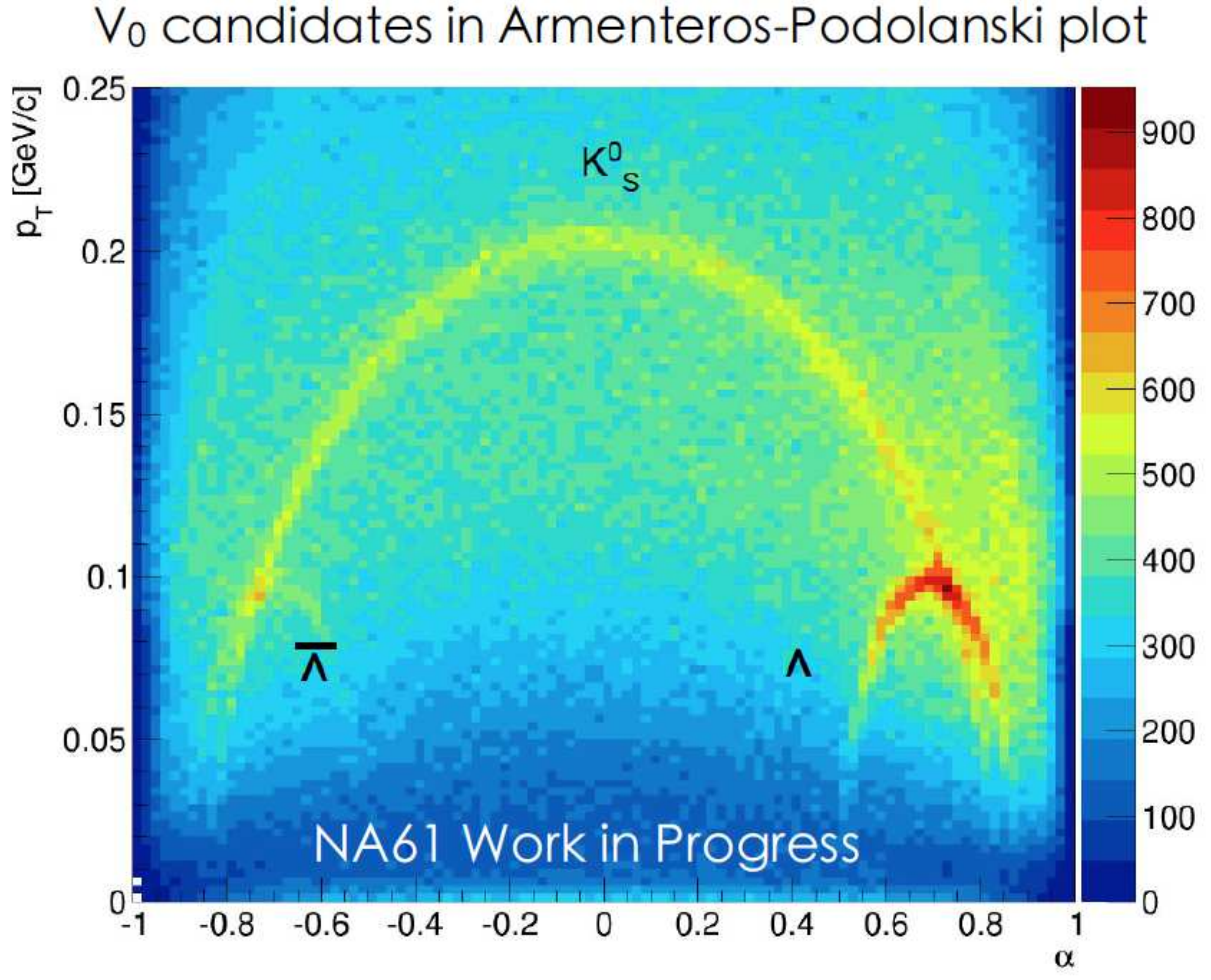}
\caption{ V$_0$ candidates found in $\pi^+$+C at 60 GeV/$c$ interactions in the 2016 data sets.
Plot with regard to transverse momentum of oppositely charged decay products with respect to the V$_0$ ($p_\mathrm{T}$) vs longitudinal momentum asymmetry ($\alpha = \frac{p_\mathrm{l}^+ - p_\mathrm{l}^-}{p_\mathrm{l}^+ + p_\mathrm{l}^-}$)
}
\label{Fig:V0candidates}
\end{figure}

In 2016, $p$+C/Be/Al  at 60 GeV/$c$, $\pi^+$+C/Be at 60 GeV/$c$, and $p$+C/Be at 120 GeV/$c$ interactions were measured.
Spectra measurements are ongoing and an example of V$_0$ candidates found in $\pi^+$+C interaction is shown in Figure~\ref{Fig:V0candidates}.
In 2017, measurements continued including FTPCs and data calibration is ongoing.
In 2018, a replica target of the NuMI beamline has been built and measurements with proton beam at 120 GeV/$c$ has been performed.

\section{Prospect: NA61 beyond 2020}

In this section, plans and possible measurements in the NA61/SHINE facility after CERN long shutdown 2 will be presented.

The next generation neutrino oscillation experiments are being proposed, such as LBNF/DUNE 
and successor experiments of T2K (T2K Phase 2 and Hyper-Kamiokande experiments).
The LBNF beamline will shoot protons with momentum somewhere between 60 and 120\,GeV/$c$ on carbon or beryllium target, similarly as the NuMI beamline with 120\,GeV/$c$ protons. 
Because primary protons have high momentum, secondary protons tend to be produced beam-forward direction and their re-interactions contribute a lot to the neutrino flux for the LBNF and NuMI beamlines. 
Therefore, it is very important to measure forward proton productions to further improve the neutrino flux prediction.
The T2K beamline will be re-used for its successor experiments with the same beam momentum and upgraded beam intensity, and there exists a possibility to re-design the target for future operations.
For all the next generation experiments, hadron production measurements with replica targets are highly desirable once their design is fixed. 


The NA61/SHINE collaboration is preparing program extension proposal after CERN Long Shutdown 2 (2019-2020). 
Significant facility modifications are planned including 
TPC electronics upgrade which significantly increases readout rate up to 1000\,Hz, 
installation of modern silicon-based tracking detectors surrounding the target which improves vertex reconstruction precision drastically,
and new time of flight detectors with improved timing resolution ($\sigma_{\rm time} \sim$50\,ps) which increases particle identification resolution, in addition to the existing large acceptance TPCs.
These plans are summarized in ~\cite{NA61Addendum}.
Therefore, it is a great opportunity for the next generation neutrino experiments to perform precise measurements of hadron production based on their demands at the upgraded NA61/SHINE facility.

\section{Summary}

Precision hadron production measurements are essential to reduce the leading systematic uncertainty on the neutrino flux prediction.
NA61/SHINE thin and replica target measurements have improved and will further improve the flux prediction in T2K.
NA61/SHINE measurements for the Fermilab neutrino programs are ongoing.
NA61/SHINE is proposing program extension after Long Shutdown 2, which significantly upgrades their facility and is a great opportunity to pursue important measurements.

\Acknowledgements
We would like to thank the CERN EP, BE and EN Departments for the strong support of NA61/SHINE.
This work was supported by the U.S. Department of Energy.


\begin{thebibliography}{99}
\bibitem{Aliaga:2016oaz} 
  [MINERvA Collaboration],
  Phys.\ Rev.\ D {\bf 94}, no. 9, 092005 (2016).
  Addendum: Phys.\ Rev.\ D {\bf 95}, no. 3, 039903 (2017).

\bibitem{Laura-NA61Beyond2020}
L.~Fields, Presentation at the NA61 Beyond 2020 Workshop.\\
\url{https://indico.cern.ch/event/629968/contributions/2659930}

\bibitem{Abgrall:2011ae} 
  [NA61/SHINE Collaboration],
  Phys.\ Rev.\ C {\bf 84}, 034604 (2011).

\bibitem{Abgrall:2011ts} 
  [NA61/SHINE Collaboration],
  Phys.\ Rev.\ C {\bf 85}, 035210 (2012).

\bibitem{Abgrall:2013wda} 
  [NA61/SHINE Collaboration],
  Phys.\ Rev.\ C {\bf 89}, no. 2, 025205 (2014).

\bibitem{Abgrall:2015hmv} 
  [NA61/SHINE Collaboration],
  Eur.\ Phys.\ J.\ C {\bf 76}, no. 2, 84 (2016).

\bibitem{Abgrall:2012pp} 
  [NA61/SHINE Collaboration],
  Nucl.\ Instrum.\ Meth.\ A {\bf 701}, 99 (2013).

\bibitem{Abgrall:2016jif} 
  [NA61/SHINE Collaboration],
  Eur.\ Phys.\ J.\ C {\bf 76}, no. 11, 617 (2016).

\bibitem{Berns:2018tap} 
  [NA61/SHINE Collaboration],
  arXiv:1808.04927 [hep-ex]. (submitted to EPJC).

\bibitem{Aduszkiewicz:2018uts} 
  [NA61/SHINE Collaboration],
  Phys.\ Rev.\ D {\bf 98}, no. 5, 052001 (2018).

\bibitem{NA61Addendum}
 [NA61/SHINE Collaboration],
\url{https://cds.cern.ch/record/2309890}.

\end{thebibliography}
\end{document}